# Star formation efficiency and flattened gradients in M31

F. Robles-Valdez[1]*,  L. Carigi[1], M. Peimbert[1]

[1] *Instituto de Astronomía, Universidad Nacional Autonóma de México, AP 70-264, 04510, México DF, México*
*E-mail: frobles@astro.unam.mx

**RESUMEN**

Presentamos un modelo de evolución química para M31 basado en un escenario de formación pronunciado. Se reproducen las tres restricciones observacionales principales del disco de M31: la masa total bariónica, la masa de gas y la abundancia de O/H. El modelo muestra buen acuerdo con las observaciones de: SFR(r), Mstars(r) y los gradientes de C/H, N/H, Mg/H, Si/H, S/H, Ar/H, Cr/H, Fe/H y Z. Para reproducir la masa de gas observada, encontramos que la eficiencia de formación estelar varía en el espacio, para el disco completo, y es constante en el tiempo la mayor parte de la evolución (t< 12.8 Gyr). Para reproducir la SFR(r) observada, encontramos que la eficiencia disminuye casi a cero para 12.8<t(Gyr)< 13.0 y r>12 kpc. Todos los gradientes Xi/H predichos muestran tres pendientes diferentes, debido a la dependencia en r de la eficiencia de formación estelar, y de la formación galáctica dentro-fuera.

**ABSTRACT**

We present a chemical evolution model for M31 based on a pronounced inside-out formation scenario. The model reproduces the three main observational constraints of the M31 disk: the radial distributions of the total baryonic mass, the gas mass, and the O/H abundance. The model shows good agreement with the observed: SFR(r), Mstars(r), the C/H, N/H, Mg/H, Si/H, S/H, Ar/H, Cr/H, Fe/H, and Z gradients. From reproducing the observed gas mass, we find that the star formation efficiency is variable in space, for the whole disk, and is constant in time for most of the evolution (t< 12.8 Gyr). From reproducing the observed SFR, we find that the efficiency decreases almost to zero for 12.8<t(Gyr)< 13.0 and r>12 kpc. All the predicted Xi/H(r) gradients show three different slopes, due to the r-dependence of the the star formation efficiency and the inside-out galactic formation.

## 1  INTRODUCTION

Chemical evolution models (CEMs) are useful to predict the chemical history of galaxies.Therefore, CEMs are able to infer, in general, how galaxies and stars have been formed and how efficient are the stars to produce chemical elements.

In this paper we present a successful CEM for the Andromeda galaxy, M31, the largest spiral galaxy in the Local Group. M31 is an Sb galaxy located at a distance of $744\pm33$ kpc (Villardell et al. 2010) and has a deprojected semi-major axis of $70\pm10$ kpc. The baryonic mass of M31 is about 1.2 times larger than that of the Milky Way (Widrow, Perrett & Suyu 2003).

There are some chemical evolution models of M31 in the literature, that were built trying to reproduce the observed radial distributions of: i) the gas mass (Mgas) without considering the He and Z contribution and ii) the O/H gaseous values of the H II regions, without consider the correction in the O/H ratio due to temperature variations and without considering the correction due to the fraction of O trapped in dust grains, these two corrections taken together reduce the true O/H value in the H II regions by about 0.3 dex (e. g. Peimbert et al. 2007, Peimbert & Peimbert 2010, Peña-Guerrero et al. 2012). The temperature variations can be represented by the mean square temperature variation $t^2$ (Peimbert 1967). Below we describe briefly the most recent CEMs of M31, for more details of these CEMs see the paper of the respective authors.

Renda et al. (2005) presented a set of CEMs for M31, based on the SFR formulation by Wyse & Silk (1989), with a star formation efficiency constant in time and space, where the SFR is inversely proportional to the radius. They used the initial mass function (IMF) by Kroupa, Tout & Gilmore (1993). Their best model reproduced reasonably well the observations available when they developed that work.

Mattsson (2008) built two CEMs of M31, assuming a star formation law that combines a Silk law and a Schmidt law (Boissier et al. 2003), with a star formation efficiency constant in time and space, and dependent of the angular frequency of the galactic disk. He used an IMF similar to Scalo's one (1986), but considered a larger fraction of very low mass stars and substellar objects. This model could not reproduce the observational radial profile of the gas mass and adjusted moderately well the observed of O/H abundances considered by him.

Yin et al. (2009) constructed a CEM for M31, considering a modified star formation Kennicutt-Schmidth law (Fu et al. 2009), with a star formation efficiency constant in time and space, and the SFR is inversely proportional to the radius. They assumed the IMF by Kroupa, Tout & Gilmore (1993). Their model did not adjust the distribution of the gas mass for r< 8 kpc. They obtained a good fit with the observed values of O/H.

Marcon-Uchida, Matteucci & Costa (2010) developed a set of CEMs of M31. They used a SFR that follows the Kennicutt-Schmidt law (Kennicutt 1998), with a star formation efficiency variable with radius, and adopting a gas mass threshold for star formation of 5 $M_\odot$ pc$^{-2}$. They assumed the IMF by Kroupa, Tout & Gilmore (1993). Their best model failed to reproduce the distribution of gas mass and presents a reasonable fit of the O/H abundances.

Carigi, García-Rojas & Meneses-Goytia (2013) built a CEM under the instantaneous recycling approximation, adopting a SFR following the Kennicutt-Schmidt law (Kennicutt 1998) with a star formation efficiency constant in time and in radius. They adopted the IMF by Kroupa, Tout & Gilmore (1993). Their CEM does not reproduce the distribution observed of the gas mass and reproduces well the most probable O/H gradient based in the observational data.

Spitoni, Matteucci & Marcon-Uchida (2013) presented a study of the effects of galactic fountains and radial gas flows on the chemical evolution models. Their study is based on the CEM by Marcon-Uchida, Matteucci & Costa (2010), described above. They concluded that galactic fountains are not an important factor in the chemical evolution of the galactic disk. Moreover they found that the radial gas flows produce the same effect on the O/H gradient that adopting a threshold and an r-dependent efficiency of the SFR.

Since the previous models cannot explain the radial distribution of the gas mass of the M31 disk, and currently there are more precise observations, we have decided to built a new CEM for M31. Our CEM reproduces successfully and for the entire disk, the updated an more precise values of the radial distributions of the total baryonic mass, the gas mass and the O/H abundances, and also shows reasonable agreements with the radial distributions of the star formation rate, the stellar mass, and the gradients of other nine chemical elements.

Throughout this paper, we adopt the usual notation *X*, *Y*, and *Z* to represent the hydrogen, helium and heavy element abundances by mass, while the abundances by number are represented without italics. In Section 2, we describe the observational constraints used in this paper. In Section 3, we present the features of our CEM. In Section 4, we present the main results of our CEM. In Section 5, we discuss the helium and oxygen enrichment. The conclusions are presented in Section 6. In the appendix we present other models that fail to reproduce one of the observational constraints.

## 2 OBSERVATIONAL CONSTRAINTS

The CEM of M31 is built to reproduce three main observational constraints of the galactic disk at the present time: the radial distributions of the total baryonic mass, the gas mass, and the O/H ratio. Also the predictions of the CEM are compared with other observational constraints of the M31 disk, the radial distributions of: the star formation rate, the stellar mass, the abundances of fifteen chemical elements relative to H and the total metallicity, *Z*.

### 2.1 Radial distribution of the mass surface densities

*2.1.1 Gas mass, Mgas(r)*

The radial profile of the gas mass, *Mgas(r)* ($M_\odot$ pc$^{-2}$), represents the atomic and molecular gas that contains all the chemical elements, *i. e. Mgas(r) = MX(r)+MY(r)+MZ(r)*. We obtain *MX(r)*, by adding the atomic and molecular components, that were provided by Tabatabaei (2013, private communication), based on Nieten et al. (2006), and we correct them by the inclination angle of 75º (Berkhuijsen 1977, Braun 1991).

We computed *MY(r)* and *MZ(r)* from the following relations:

i) The proportion between gas mass of a given element, *Mi*, and the chemical abundance by mass of this element, *Xi*, is *Mi = Xi x Mgas*. Specifically: *MY(r) / MX(r) = Y(r) / X(r)* and *MZ(r) / MX(r) = Z(r) / X(r)*.

ii) The normalization of the chemical abundances by mass: *X(r)+Y(r)+ Z(r)= 1*.

iii) The *Y(r)* and *O(r)* enrichment by Carigi & Peimbert (2008): *Y(r) = 0.2477+3.3O(r)*.

iv) The transformation of *O/X* (by mass) to O/H (by number): *O(r)/X(r) = 16* (O/H)(r).

v) The O/H gradient determined in this paper (see eq. 4).

vi) The conservation of the solar Z / (O/H) ratio in the interstellar medium: *Z(r) / (O/H)(r) = $Z_\odot$ / (O/H)$_\odot$*.

vii) The protosolar values of $Z_\odot = 0.0142$ and $12 + \log(O/H)_\odot = 8.73$ dex by Asplund et al. (2009).

The contributions of *MY* and *MZ* increase the gas mass surface density approximately by a factor of 1.2, relative to *MX*. It is worthy to note that the radial distribution of the corrected gas mass shows a broad peak between 9 and 11 kpc (see Figure 1, top panel) that is not present in the MW (see Esteban et al. 2013).

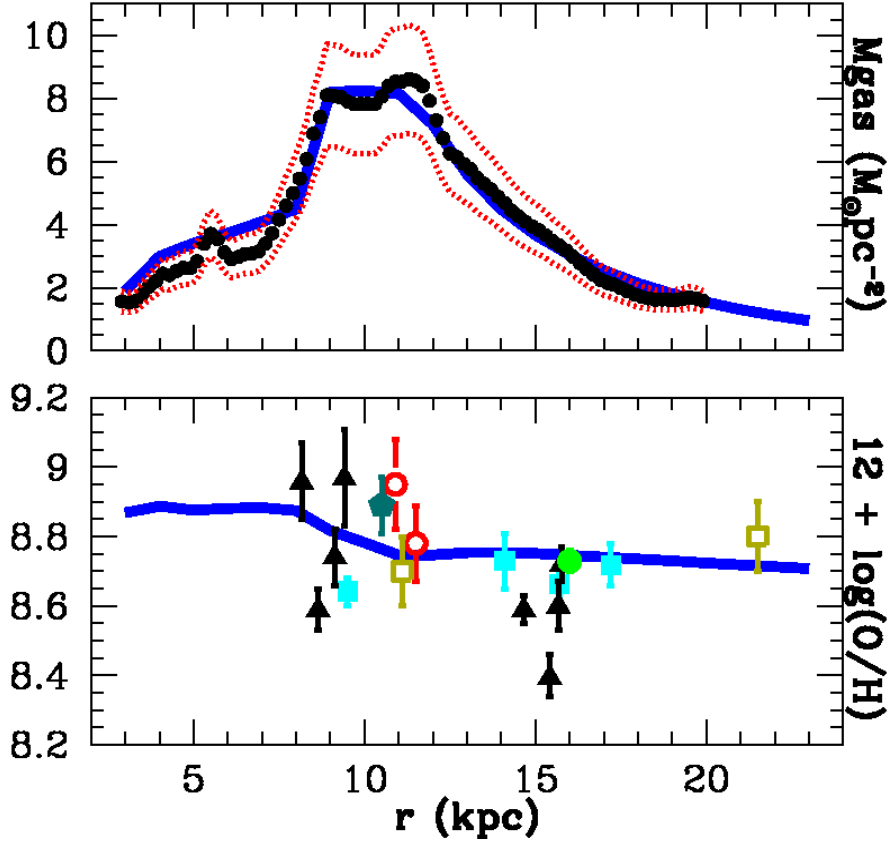

Figure 1. Top panel: radial distribution of the gas mass at the present time of our chemical evolution model (CEM, solid blue line). Observational data by Tabatabaei & Berkhuijsen (2010) plus contributions of helium and metals (black circles, see Section 2.1.1). The red dotted lines represent the values within one σ error. Bottom panel: radial distribution of the O/H ratios at the present time, CEM (solid blue line). Observational data: (i) gaseous values from H II regions corrected by us for temperature variations ($t^2$ factor) and dust depletion (see Section 2.3), Toribio-San Cipriano et al. (2014) (filled darkgreen pentagon), Zurita & Bresolin (2012) (filled black triangles), Sanders et al. (2012) (filled cyan squares), and Esteban et al. (2009) (filled green circle); (ii) stellar values: Przybilla et al. (2006) (empty red circles), and Venn et al. (2000) (empty gold squares). The vertical lines represent one σ error.

*2.1.2 Stellar mass, Mstars(r)*

We use the luminosity profile of the M31 disk to obtain the radial distribution of the mass surface density of living stars, $MstarL(r)(M_\odot\ pc^{-2})$. We consider an average of the disk scale lengths from three bands: R, K, and I by Walterbos & Kennicutt (1988), Hiromoto et al. (1983), and Worthey et al. (2005), respectively, and we obtain a disk scale length equal to rdisk=5.5 kpc. We calculate the central density by the integration of the $MstarL(r))(M_\odot\ pc^{-2})$ over the surface of the galactic disk in order to reproduce the stellar mass estimated for the M31 disk, which is 6 x $10^{10} M_\odot$ (Tamm et al. 2012).

In addition, we take into account the contribution to the stellar mass due to the stellar remnants, MstarR(r))(M$_\odot$ pc$^{-2}$), with a value equal to 13 percent of the living stellar mass, according to the computation for the Galaxy (based on the Milky Way model computed by Carigi & Peimbert (2011). Therefore, the radial distribution of the total stellar profile is given by:

$$\text{Mstars}(r) = \text{MstarL}(r) + \text{MstarR}(r) = 230\ e^{(-r/5.5\ \text{kpc})}, \quad (1)$$

see Figure 2, top panel.

*2.1.3 Baryonic mass, Mtot(r)*

The total baryonic mass is the sum of the profiles of the gas mass and the total stellar mass, *i. e.* Mtot(r) = Mgas(r) + Mstar(r). Therefore, the radial distribution of the total baryonic mass shows a single exponential profile, given by:

$$\text{Mtot}(r) = 240 e^{(-r/5.7\text{kpc})}. \quad (2)$$

**2.2. Radial distribution of the star formation rate, SFR(r)**

M31 is a galaxy with a relatively quiescent star formation at the present time, considering that the values estimated of its total star formation rate are $\sim$ 0.35−1.0 M$\odot$ yr$-1$ (see table 1, Yin et al. 2009). This range is smaller than the values estimated for the Milky Way, that amount to 1.5-2.3 M$\odot$ yr-1 (Kennicutt & Evans 2012).

To compare with our CEM, we use the radial distribution of the star formation rate obtained by Boissier et al. (2007), estimated from UV profiles, corrected for extinction by dust (see Figure 2, bottom panel). The total star formation rate using the data by Boissier et al. is about 0.8 M$\odot$ yr-1. The observations show that the SFR(r) and Mgas(r) do not obey a Kennicutt-Schmidt law with a constant v, due to a decrease in the observed values of the SFR.

**2.3. Radial distribution of the chemical abundance ratios, Xi (r)/H(r)**

As observational constraints at the present time, we consider the most recent studies on the chemical abundances of H II regions and of supergiant stars, because both represent the current chemistry of the gas in the M31 disk.

To obtain the O/H abundances of H II regions we take into account the recalibration method called Corrected Auroral Line Method (CALM), presented by Peña-Guerrero, Peimbert & Peimbert (2012). In this calibration method, they considered the temperature inhomogeneity in H II regions given by the t$^2$ factor (Peimbert 1967; Peimbert et al. 2007a), and the fraction of oxygen atoms trapped into dust grains (Peimbert & Peimbert 2010). We use their equation to obtain the corrected oxygen abundance given by:

$$(12 + \log(\text{O/H})_{\text{CALM}}) = 1.0825\ (12 + \log(\text{O/H})_{4363/5007}) - 0.375, \quad (3)$$

where (O/H)4363/5007 correspond to the abundance estimated with the temperature derived from the [O III] (4363/5007). Consequently, we use only the chemical abundances estimated from the brightest H II regions, where the auroral lines can be observed. Recently Croxall et al. (2013) have determined O/H ratios for the galaxy NGC 628 based on the infrared lines of [O III]. These lines do not depend on the temperature structure of the H II region and the abundances are in good agreement with the CALM calibration by Peña-Guerrero et al. (2012).

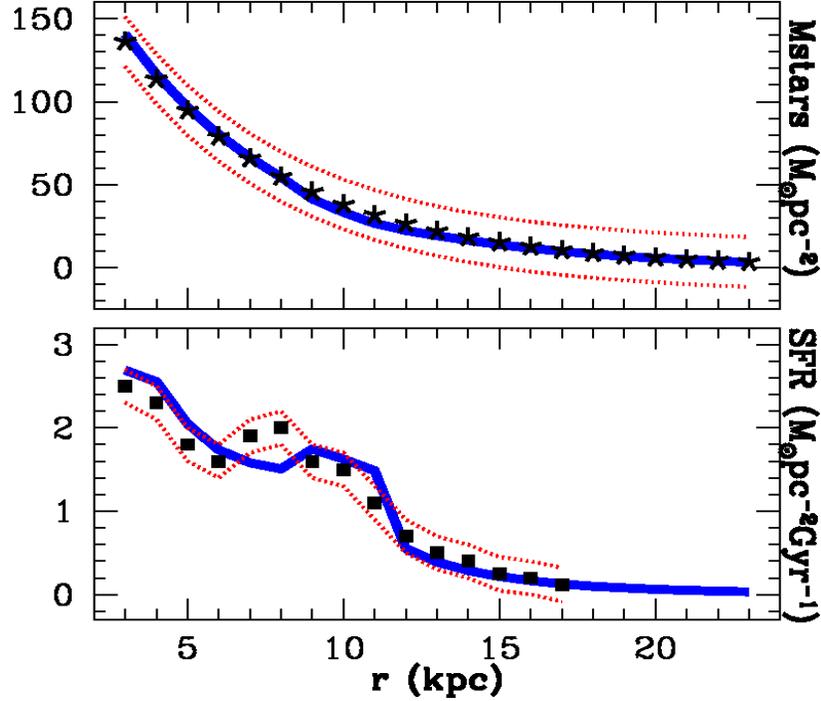

Fig. 2. Top panel: radial distribution of the total stellar mass at the present time, CEM (solid blue line). Observational data: an average of the disk scale length equal to 5.5 kpc, and a central density equal to 230 M⊙ pc$^{-2}$ (black stars, see Section 2.1.2). Bottom panel: radial distribution of the SF R at the present time, CEM (solid blue line). Observational data: Boissier et al. (2007) (filled black squares, see Section 2.2). The red dotted lines represent the values within one σ error.

Toribio-San Cipriano et al. (2014, in prep.) derived the C/H and O/H abundances in one H II region, K160, using collisional excitation lines (CELs). Since CELs are sensitive to spatial temperature variations in the H II regions, and considering the dust-phase component of the oxygen, they corrected the O/H ratio with equation 3 shown in our article, and the C/H ratio by dust depletion, the dust depletion increases the gaseous values by 0.10 (E09).

Zurita & Bresolin (2012) derived the N/H, O/H, S/H, Ar/H, and Ne/H abundances for eight H II regions using collisional excitation lines (CELs), and assuming the [O III] (4363/5007) temperature. Since CELs are sensitive to spatial temperature variations in the H II regions, and considering the dust-phase component of the oxygen, their O/H ratios are corrected using equation (3). In the case of the N/H, S/H, Ar/H, and Ne/H abundances we correct by the t$^2$ factor, increasing the gaseous values by 0.19, 0.21, 0.24, and 0.26 dex, respectively (Esteban et al. 2009, E09). The correction due to the fraction of N and S embedded in dust has not been estimated for H II regions and is not taken into account; while for Ar and Ne, the correction is zero, because they are noble gases and are not expected to be embedded in dust grains.

Sanders et al. (2012) determined the N/H and O/H ratios for hundreds of H II regions from CELs, but only for four H II regions they used the [O III] (4363/5007) temperature. We correct the O/H gaseous values for these four regions using equation 3. The N/H values are increased by 0.26 dex, due to the t2 factor (E09), and no correction is made for the fraction of N atoms embedded in dust grains.

Esteban et al. 2009 obtained the C/H, N/H, O/H, Ne/H, S/H, Ar/H, and Fe/H abundances in one H II region: K932, using recombination lines (RLs) and the [O III] (4363/5007) temperature. The RL ratios are not sensitive to temperature variations, therefore, the abundances based on RL ratios are not corrected by the $t^2$ factor. We correct the C/H, O/H and Fe/H abundances by dust depletion, increasing the gaseous values by 0.10, 0.11 and 1.00 dex, respectively (E09; Mesa-Delgado et al. 2009; Peimbert & Peimbert 2010). The N/H, Ne/H, and Ar/H abundances by E09 are not increased by dust depletion.

Considering the corrected O/H values from H II regions and their estimated error, we obtain the oxygen gradient in the 7-18 kpc region of the M31 disk:

$$12 + \log(O/H) = (8.73 \pm 0.06) - (r - 12)(0.013 \pm 0.006). \qquad (4)$$

We compute the Z value of each H II region, adding all chemical abundances and dividing by the sum of the protosolar abundances (Asplund et al. 2009) of the same elements considered for each H II region, thus:

$$Z_{HII} = (\Sigma(X_i)_{HII} / \Sigma(X_i)_{\odot}) Z_{\odot}, \qquad (5)$$

(see Figure 7, bottom panel).

There are several chemical abundance determinations for O, B, A and F supergiant stars in M31 that are used by us as a consistency check for the predicted O/H gradient of our CEM. Przybilla et al. (2006) derived the C/H, N/H, O/H, Mg/H, Al/H, Si/H, S/H, Sc/H, Ti/H, Cr/H, Mn/H, and Fe/H ratios for two A stars. Venn et al. (2000) determined the C/H, O/H, Mg/H, Si/H, Ca/H, Sc/H, Ti/H, Cr/H, Mn/H, Fe/H and Ni/H ratios for four A-F stars.

The gaseous O/H values become similar to the stellar ones after the gaseous values have been increased by ~ 0.3 dex. This increase is due to two corrections, one by the $t^2$ factor and the another by the fraction of O atoms trapped in dust. Both corrections have been considered in equation 3.

We calculate also the Z values of the supergiant stars, in the same way as for the H II regions, thus:

$$Z_{stars} = (\Sigma(X_i)_{stars} / \Sigma(X_i)_{\odot}) Z_{\odot}, \qquad (6)$$

(see Figure 7, bottom panel).

Figure 1, bottom panel, shows the O/H ratios from H II regions and from stars, and Figures 4 to 7 show the chemical abundances for the other fifteen chemical elements and for Z.

We used the Si/H, Ca/H, Ti/H, and Fe/H ratios of globular clusters of M31 estimated by Colucci et al (2009), in order to constrain the efficiency of the star formation rate in the halo.

## 3. CHEMICAL EVOLUTION MODEL, CEM

In this work, we present a chemical evolution model for the M31 disk, based on the standard chemical evolution equations, which were originally written by Tinsley (1980) and are currently used by several authors (e.g. Pagel 2009, Cescutti et al. 2012, Carigi & Peimbert 2011, and Prantzos 2012). The model is built to reproduce the present radial distributions of: the total baryonic mass, the gas mass, and the O/H values.

The characteristics of the model are:

**i)** The halo and disk are projected onto a single disk of negligible width and with azimuthal symmetry; therefore, all the functions are dependent only on the galactocentric distance r and time t. The single disk consists of a series of concentric rings without exchange of matter between them ranging from 3 to 23 kpc and each ring has a width of 1 kpc.
**ii)** The model describes the r ≥ 3 kpc region, because the physical processes associated with the bulge and bar in the central region of M31 are not considered.
**iii)** The age of the model is 13.0 Gyr, the time elapsed since the beginning of the formation of the galaxy to the present time.
**iv)** The model is built based on an inside-out scenario, like is predicted by cosmological models (Avila-Reese, Firmani, & Vázquez-Semadeni 2003) with a double infall of primordial abundances: $Y_p = 0.2477$, and $Z = 0.00$ (Peimbert et al. 2007). The double-infall rate adopted is similar to that presented by Chiappini, Matteucci & Gratton (1997), is a function of r and t: $IR(r, t) = A(r)e^{-t/\tau_{halo}} + B(r)e^{-(t-1Gyr)/\tau_{disk}}$. The halo formation occurs during the first Gyr with a timescale formation $\tau_{halo} = 0.1$ Gyr, and the disk formation takes place from 1 Gyr until 13 Gyr, with a timescale that depends on the radius $\tau_{disk} = 0.3r(kpc) - 0.2$ Gyr. To obtain the A(r) and B(r), we adopt the mass density of the disk and the halo components found for the Milky Way by Carigi & Peimbert (2011), $M_{halo}/M_{disk} = 0.25$. Then, we chose the variables $A(r) = 48\, e^{-(r/5.7\, kpc)}$ and $B(r) = 192\, e^{-(r/5.7\, kpc)}$, in order to reproduce the present radial distribution of the total baryonic mass in the M31 disk.
**v)** The model does not consider: a) the loss of gas and stars from the galaxy to the intergalactic medium, and b) the gaseous and stellar radial flows through the galactic disk.
**vi)** The star formation rate SFR was parametrized as the Kennicutt-Schmidt Law (Kennicutt 1998): $SFR(r, t) = \nu(r) M_{gas}(r, t)^n$, with $n = 1.4$. Here $\nu$ is the star formation efficiency obtained by adjusting the current gas mass radial distribution in the galactic disk. Moreover, during the first Gyr we assumed a $\nu$ value 5 times higher during the halo formation than that adopted for the disk, in order to match mainly the Fe/H abundances shown in globular clusters (see section 2.3).
**vii)** The model uses the initial mass function (IMF) proposed by Kroupa, Tout & Gilmore (1993), in the mass interval given by 0.08 M⊙ to an upper mass, $m_{up}$, which is chosen to be 45 M⊙ to reproduce the current absolute value of the O/H gradient of the M31 disk.
**viii)** Stars enrich the interstellar medium according to a matrix computed by Robles-Valdez, Carigi & Peimbert (2013) from metal-dependent yields of massive stars (MS; $8 < m/M_\odot < m_{up}$), low and intermediate mass stars (LIMS; $0.08 < m/M_\odot < 8$), and Type Ia SNe. The set of yields includes:
a) For stars with masses from 13 to 40 M⊙ and with $Z = 0.0, 0.001, 0.004$, and $0.02$, we adopt the yields by Kobayashi et al. (2006). In the cases of $Z = 0.008$ and $0.02$, we add to the yields by Kobayashi et al. an average of the stellar winds yields of intermediate mass loss rate for He, C, N and O, obtained by Maeder (1992) and Hirshi, Meynet & Maeder (2005), following the suggestion by Carigi & Peimbert (2011). For $Z = 0$, we increase the yields by Kobayashi et al. adding to them the rotation yields for He, C, N, and O, estimated by Hirshi (2007), that modification is based on the results by Kobayashi, Karakas & Umeda (2011).
b) For star with masses from 1 to 6 ⊙ and with *Z*= 0.0001, 0.004, 0.008 and 0.02, we use the yields by Karakas (2010).
c) For binary stars, with total mass between 3 and 16 M⊙, we adopt the yields by Nomoto et al. (1997) in the Ia SNe formulation by Greggio & Renzini (1983). We use the variable Abin = 0.05 as the fraction of binary stars that are progenitors of IA SNe. This Abin value is chosen in order to reproduce the current Fe/H gradient.

**4. RESULTS AND DISCUSSION**

In this section we present the results of the best model. This model reproduces all the main observational constraints of the M31 disk. Before obtaining our best model, we built other models considering different combinations of the input parameters. Nevertheless these models were not successful to reproduce the observations for all radii at t= 13 Gyr. We show some of these unsuccessful model in the appendix.

As Figure 1 shows, our best chemical evolution model reproduces well Mgas(r) and O/H(r) at the present time and for all galactocentric distances. Since for M31 the number of model free parameters and the number of the reliable observational constraints are similar, the model can match the main observational constraints with only one reasonable combination of the CEM ingredients. Therefore, given an infall, the behavior of the Mgas depends mainly on the chosen SFR, and consequently in this work, on the efficiency of star formation for each r, ν(r). When a chemical yield set is assumed, the slope of the O/H gradient is a consequence of the adopted SFR and the absolute value of this gradient is determined by the mup of the IMF. For the model it was necessary to adopt a pronounced inside-out scenario of galactic formation (inner parts are formed much faster than the outer parts) in order to reproduce the observed low Mgas for r< 8 kpc.

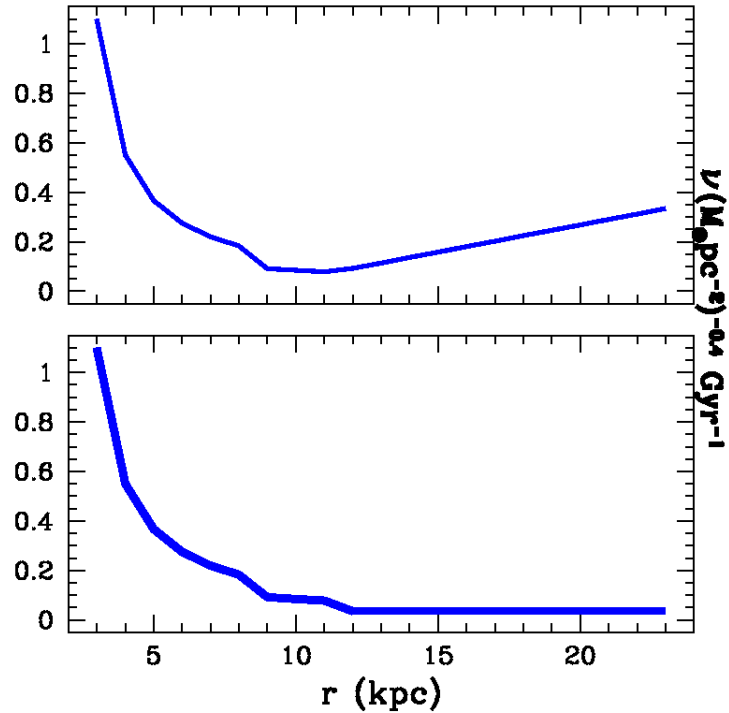

Fig. 3. Radial distribution of the star formation efficiency, ν = SFR/Mgas$^{1.4}$, of our CEM. Top panel: for 0 < t(Gyr)≤ 12.8 (thin blue line). Bottom panel: for 12.8 < t(Gyr)≤ 13.0 (thick blue line).

To construct our successful model, we built a CEM with a ν variable in space to reproduce the current Mgas for all radii. In Figure 3 (upper panel) we show the radial distribution of the star formation efficiency, for 0.0< t(Gyr)< 12.8, which follows the equations: ν ∼ 4.32/r − 0.34 for the inner disk, ν ∼ −0.04r + 0.55 for the middle disk, and ν ∼ 0.02r − 0.15 for the outer disk. The r-behavior of the SFR efficiency can be caused by various physical conditions in the gas of the M31 disk, e. g.: distribution and temperature of the molecular gas and turbulence processes (Leroy et al 2008, Zamora-Avilés & Semadeni 2013). As is well known, the cause, and therefore the evolution of the SFR is still an open problem in astronomy.

With these ν(r) values (constant in time) the model does not matches the estimated SFR for r > 12 kpc, because the observed SFR decreases considerably in this outer zone. Therefore, the model requieres to diminish ν(r) to a near zero value, only for r > 12 kpc and at very recent times, to keep the agreement with Mgas (r) and O/H(r). We decided to reduce ν(r) for 12.8< t(Gyr)< 13.0 (see Figure 3, lower panel) based on the results by Block et al. (2006). They, using numerical simulations, explain the ring-like structure observed in the M31 disk at ∼10 kpc as due to a 0.2 Gyr-ago encounter with a companion galaxy (probably M32), . This encounter reduced the gas mass density in the outer regions of M31, and consequently a reduction of the star formation efficiency. With this ν reduction, the model reproduces the outer observed SFR and keeps the excellent agreement with other observational constraints (see Figures 1 and 2).

In Figures 4-7, we compare our model with the best observations of the chemical abundances. There is a good agreement with the gradients of: C/H, N/H, Mg/H, Si/H, S/H, Ar/H, Cr/H, Fe/H and Z. On the other hand, our model does not show a good agreement with the gradients of: Ne/H, Sc/H, Ti/H, Mn/H, and Ni/H. Kobayashi et al. (2006, 2011), using their own yields, did not obtain good agreement for Sc/Fe, Ti/Fe, Mn/Fe, and Ni/Fe ratios in the solar vicinity.For Ne/Fe, they could not compare their model results, due to the lack of observational data. It is important to mention that Kobayashi et al. took into account only stellar abundances as observational constraints and that the line intensities of those elements are very weak in the stellar spectrum, hindering the determination of those abundances.

If we multiply the original Ne, Al, Sc, Ti, Mn, and Ni yields by Kobayashi et al. (2006) by 0.35, 2.5, 10.0, 5.0, 8.0, and 5.0, respectively, the theoretical gradients become similar to the observed ones. Previous authors applied corrections to the original yields computed by Kobayashi et al., e.g.: Hernández-Martínez et al. (2011) estimated that the original Cl and Ar yields should be multiplied by 6 and 2, respectively, to adjust the Cl/H and Ar/H gradients from H II regions, and young and old planetary nebulae in the dwarf irregular galaxy NGC 6822. Cescutti et al. (2012) multiplied the original P yield by 3 to match the P/H gradient from stars in the Milky Way disk. Robles-Valdez et al. (2013) multiplied the original Ar yield by 2 to reproduce the Ar/H gradient from H II regions in the M33 disk.

All the predicted Xi/H gradients, at the present time, show three slopes (see Figures 1 and 4-7): slightly positive for 3 < r(kpc)< 8 (inner disk), negative for 8 < r(kpc)< 11 (central disk), and slightly negative for 11 <r(kpc)< 23 (outer disk). The differences among gradients are due to the radial variations of ν. In these regions ν(r) presents three different behaviors for the most of evolution (see Figure 3, upper panel).

Radial flattenings in the innermost and outermost parts have been observed in spiral galaxies (Vlajic 2010; Goddard et al. 2011; Bresolin, Kennicutt & Ryan-Weber 2012, Rosales-Ortega et al. 2011; Sánchez et al. 2013). Esteban et al. (2013, and references therein) have observed the flattening in the outer disk of the Milky Way, this behaviour is reproduced by their CEM. They increased the ν values for r> 10 kpc ($R_{25}$MW), in order to reproduce the outer flattened gradients in the MW. The increase of ν did not affect the agreement with other observational constraints of the MW. Similarly in our CEM, ν is increased for r> 12 kpc (1/2 $R_{25}$M31), in order to reproduce the Mgas, and with the adjustments,the model fits very well the chemical abundances in the outermost regions.To confirm that the chemical gradients in the inner and outer parts of the Andromeda galaxy are flat, we need more observations.

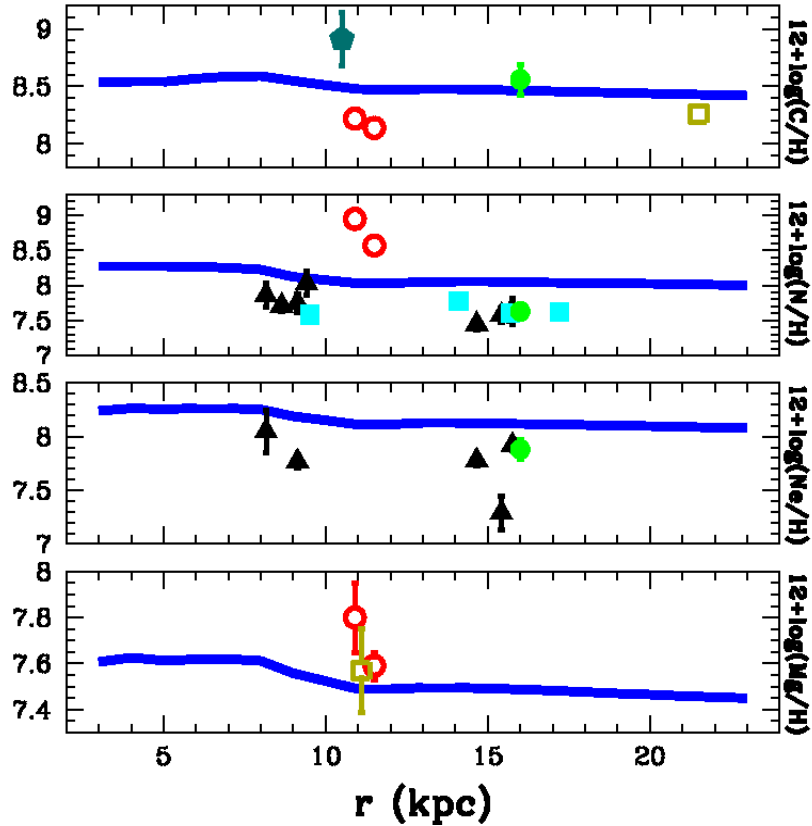

Fig. 4. Radial distributions at the present time of the C/H, N/H, Ne/H, and Mg/H ratios, CEM (solid blue line). Observational data: (i) gaseous values from H II regions corrected by us for temperature variations ($t^2$ factor) and dust depletion (see Section 2.3), Toribio-San Cipriano et al. (2014) (filled darkgreen pentagon), Zurita & Bresolin (2012) (filled black triangles), Sanders et al. (2012) (filled cyan squares), and Esteban et al. (2009) (filled green circle); (ii) stellar values, Przybilla et al. (2006) (empty red circles), and Venn et al. (2000) (empty gold squares). The vertical lines represent one σ errors. Symbols without vertical lines mean abundance ratios without computed errors.

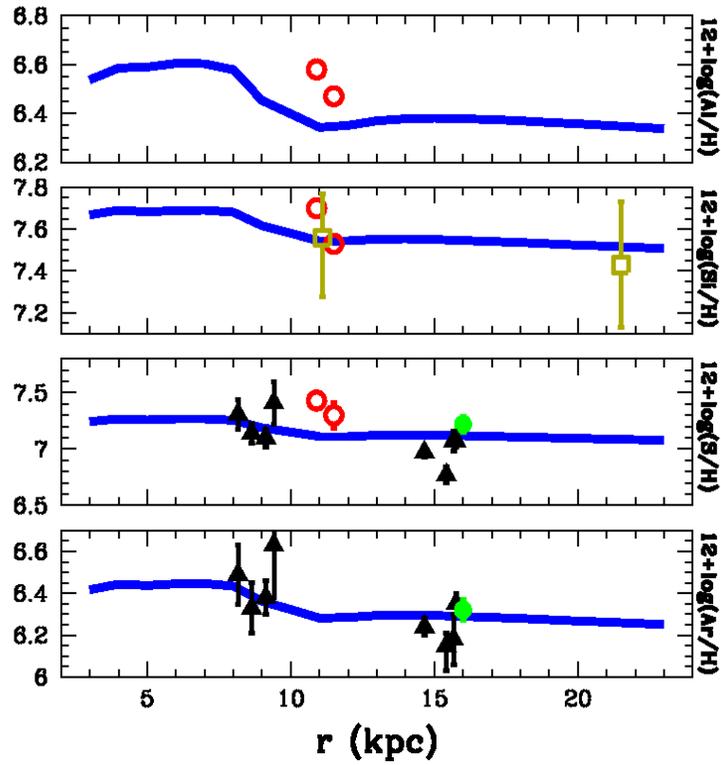

Fig. 5. Radial distributions at the present time of the Al/H, Si/H, S/H, and Ar/H ratios, CEM (solid blue line). Observational data: symbols and errors as in Fig 4.

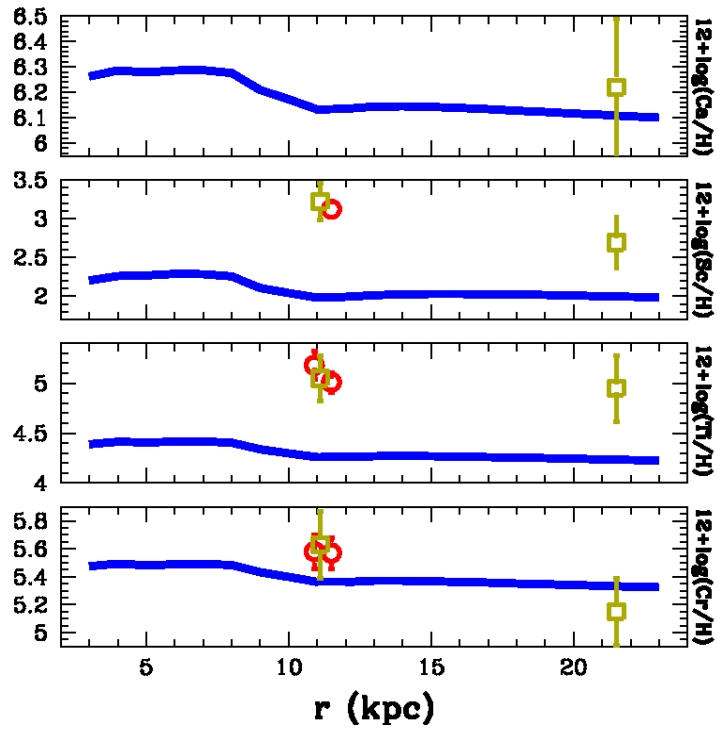

Fig. 6. Radial distributions at the present time of the Ca/H, Sc/H, Ti/H, and Cr/H ratios, CEM (solid blue line). Observational data: symbols and errors as in Fig 4.

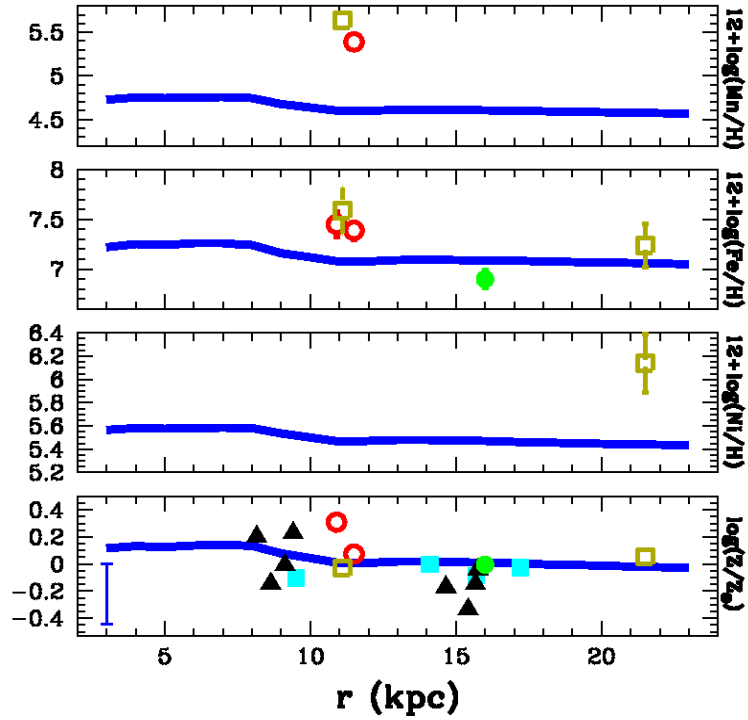

Fig. 7. Radial distributions at the present time of the Mn/H, Fe/H, Ni/H, and $Z/Z_\odot$ ratios, CEM (solid blue line). Observational data: symbols and errors as in Fig 4. Only in the bottom panel the vertical blue line in the left corner represents an average error.

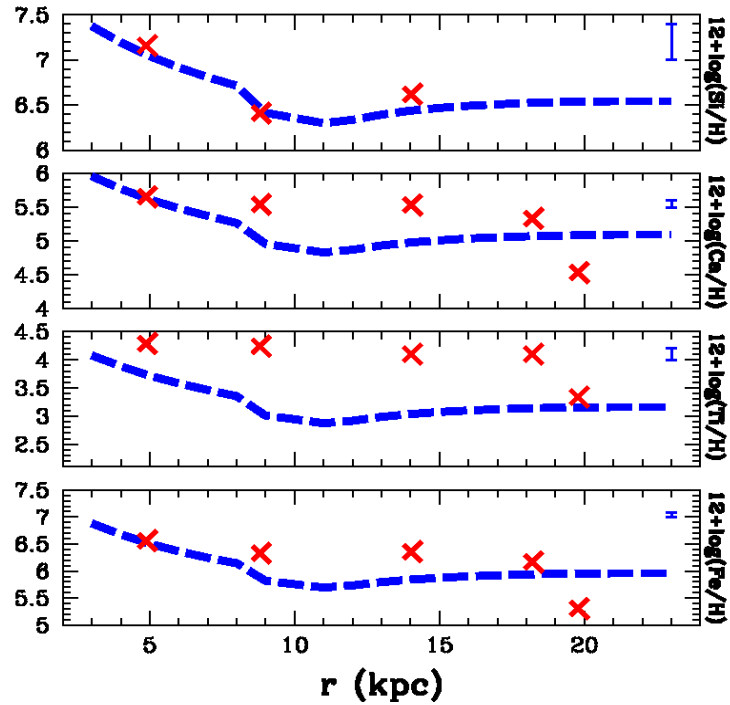

Fig. 8. Radial distributions at the present time of the Si/H, Ca/H, Ti/H, and Fe/H ratios predicted by CEM (dashed-blue line) when the halo formation is completed (1 Gyr). Observational data: from globular clusters by Colucci et al (2009) (red crosses). The vertical blue lines represent average errors.

In Figure 8, we plot the gradients of Si/H, Ca/H, Ti/H, and Fe/H at 1 Gyr, after the halo formation. Moreover we compare these gradients with the gradients observed from the globular clusters of M31 (Colucci et al. 2009), showing a reasonable adjustment, with the exception of titanium. The average disagreement (1.5 dex) of the Ti/H valuescan be seen in the halo gradient (1 Gyr, Figure 8) and in the disk gradient (13.0 Gyr, Figure 6). Therefore, the Ti yields for massive stars might have been underestimated.

In Figure 9, we show the radial distributions of: Mgas , SFR, Mstars , and O/H abundances at four times: 1, 4, 8, and 13 Gyr. Only for the SFR plot, we add the predicted values at 12.8 Gyr (see thin blue line) to compare with the 13-Gyr values, in order to show the effect of the ν reduction during the last 0.2 Gyr. Between 1 and 4 Gyr, Mgas increases due to the enormous amount of primordial gas that falls to form the disk, and the SFR increases as Mgas because the SFR is proportional to the $M_{gas}^{1.4}$. Between 4 and 13 Gyr, and mostly in the very inner radii, the decline of the infall causes the decrease of Mgas and consequently the reduction of the SFR, although the ν is higher. Since the Mstars is a time-accumulative process of the SFR, Mstars always increases with time. The O/H(r) values evolve as the inside-out scenario sets down (Carigi 1996). The predicted flattenings in the O/H gradient are due mainly to the r-behavior of ν.

In the lower panel of Figure 9, the radial distribution of O/H shows three gradients that flatten at different rates, during the evolution. The differences in the flattening rates are due to the changes in the O enrichment and the H dilution. The O enrichment depends on the SF R through the adopted ν values, and the H dilution depends on the infall rate predicted by the inside-out scenario. For r < 8 kpc, the original negative gradient, produced by a higher SFR to infall ratio in the inner radii, becomes slightly positive, because the SFR is more important at the central radii than at the inner radii during the last half of the evolution. For 8 < r (kpc) < 11 kpc, the flattening with time is less pronounced than that for the inner region, because ν is not so r-dependent, and the gradient evolution is dominated by the infall rate. For > 11 kpc, the original positive gradient, produced by a higher SFR to infall rate in the outer radii, becomes slightly negative, due to the gas dilution in the outer radii at recent times.

In Figure 10, we present the evolution of: Mgas , SFR, Mstars, and O/H values at four radii: 3, 8, 13 and 23 kpc. For r = 3 kpc and at early times, the values of these properties are higher than those for outer radii, due to the large quantity of material accreted into the inner radii compared to the outermost parts (inside-out scenario, see Figure 10, right panel). For r = 3kpc and t > 10 Gyr, Mgas increases due to the bulk of very-low-mass stars (VLMS) that formed during the first 3 Gyrs and that died recently, ejecting as a whole, huge amounts of gas to the interstellar medium. Since the expelled material by these VLMS is composed mainly by hydrogen and helium, the dying stars dilute the interstellar medium in heavy elements, and consequently O/H decreases. For the outermost radii and t > 12.8 Gyr, the SFR diminishes due to the reduction of the star formation efficiency, necessary to reproduce the observed SFR. As mentioned before, Mstars increases with time for all radii (see Figure 9), but that increase is different for each radii at a given time (see Figure 10). For inner radii, the increase rate of Mstars is higher at early times, on the contrary for outer radii, the increase rate is higher at recent times. This behavior is consequence of the inside-out scenario.

In the left panels of Figure 11 we present the radial distribution of infallat eight times: 1, 2, 3, and 4 Gyr (left-top panel), and 6,8, 11, and 13 Gyr (left-bottom panel). In the right panel,we show the evolution of infall at four radii: 3, 8, 13,and 23 kpc. It can be noted that this infall rate producesa more pronounced inside-out galactic formation than that of theMilky Way (Carigi & Peimbert 2011) and that of M33 (Robles-Valdez et al. 2013). At early times, the accretion in our M31model is much more intense in the inner regions; but atrecent times, the accretion is more important in the outerregions compared to the infall rate inferred for the MW and M33.

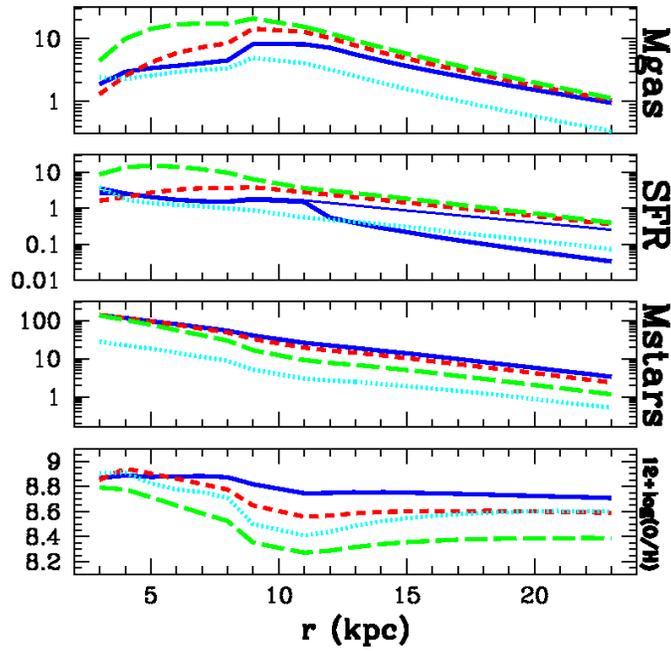

Fig. 9. Radial distributions of: Mgas (M⊙ pc$^{-2}$), SF R (M⊙ pc$^{-2}$ Gyr$^{-1}$), Mstars (M⊙ pc$^{-2}$), and O/H abundances for our CEM at four times: 1, 4, 8 and 13 Gyr (dotted cyan, long-dashed green, short-dashed red and solid blue lines, respectively). Only for the SF R the r-behavior at t = 12.8 Gyr is shown (thin blue line).

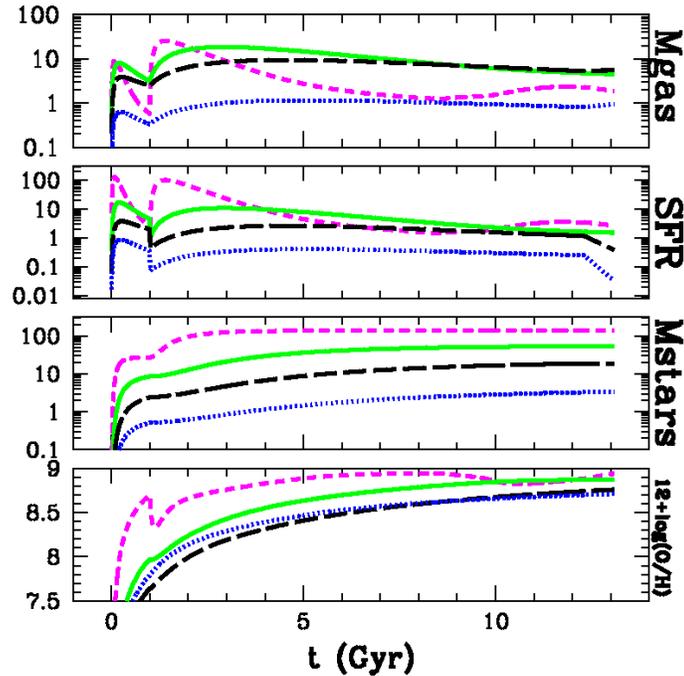

Fig. 10. Evolution of: Mgas , SF R, Mstars , and O/H abundances for our CEM at fourradii: 3, 8, 13, and 23 kpc (short-dashed magenta, solid green, long-dashed black, and dotted blue lines, respectively).

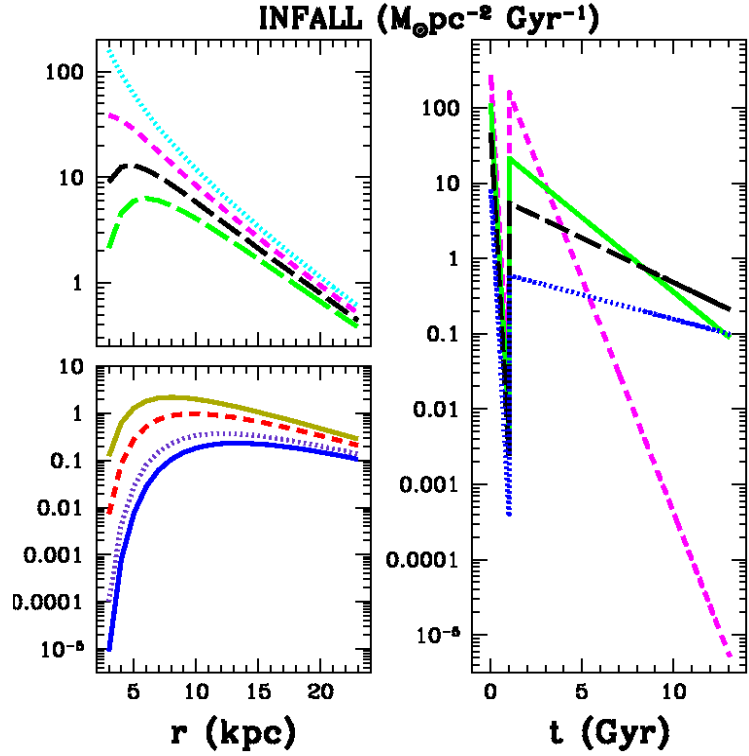

Fig. 11. Infall rate for our CEM. Left top panel: radial distribution at four times:1, 2, 3 and 4 Gyr (dotted cyan, short-dashed magenta, long-dashed black and long-dashed green lines, respectively). Left bottom panel: radial distribution at four times: 6, 8, 11 and 13 Gyr (solid gold, short-dashed red, dotted violet, and solid blue lines, respectively). Right panel: Evolution at four radii: 3, 8, 13, and 23 kpc (short-dashed magenta, solid green, long-dashed black, and dotted blue lines, respectively).

## 5. THE HELIUM AND OXYGEN ENRICHMENT IN M31

$\Delta Y$ is the helium mass fraction produced after the primordial nucleosyntesis and it is obtained from the $Y$ value of an H II region minus the primordial helium abundance, $Y_p$. On the other hand, $\Delta O$ is the oxygen mass fraction in a given H II region.

In Figure 12, we present the evolution of $\Delta Y$ vs $\Delta O$ and $\Delta Y / \Delta O$ ratio obtained from our CEM for r = 15 kpc. In general the values of $\Delta Y$ and $\Delta O$ increase with the evolution, but in the Figure 12 (top panel) presents a decrease in the values, caused by the dilution of the interstellar medium, due to the huge amount of primordial gas accreted at the beginning of the second infall. In the bottom panel, the model shows an important increase of the $\Delta Y/\Delta O$ ratio for t< 1.5 Gyr, because the most massive stars enrich the interstellar medium with more helium than oxygen, then the LIMS enrich with helium but not with oxygen.

Moreover, we compare the theoretical values at 13 Gyr with observed values for r = 14.7 kpc. $\Delta Y$ is the helium mass fraction produced after the primordial nucleosyntesis and it is obtained from the Y value of an H II region minus the primordial helium abundance, $Y_p$. On the other hand, $\Delta O$ is the oxygen mass fraction in a given H II region. The study of the $\Delta Y /\Delta O$ ratio is important to check the consistency of the chemical evolution model since He is produced by LIMS and MS, while O is mainly produced by MS. For this reason the $\Delta Y /\Delta O$ ratio depends on: i) the IMF, because it gives the ratio between the number of LIMS and MS; ii) the SFR, because it provides the total stellar mass; iii) the stellar lifetimes, because they are equivalent to the time delays between the star formation and the contribution of LIMS and MS to the ISM enrichment; iv) the He and O yields, because they represent the stellar efficiency of producing He and O; and v) gas flows in and out of the considered volumes.

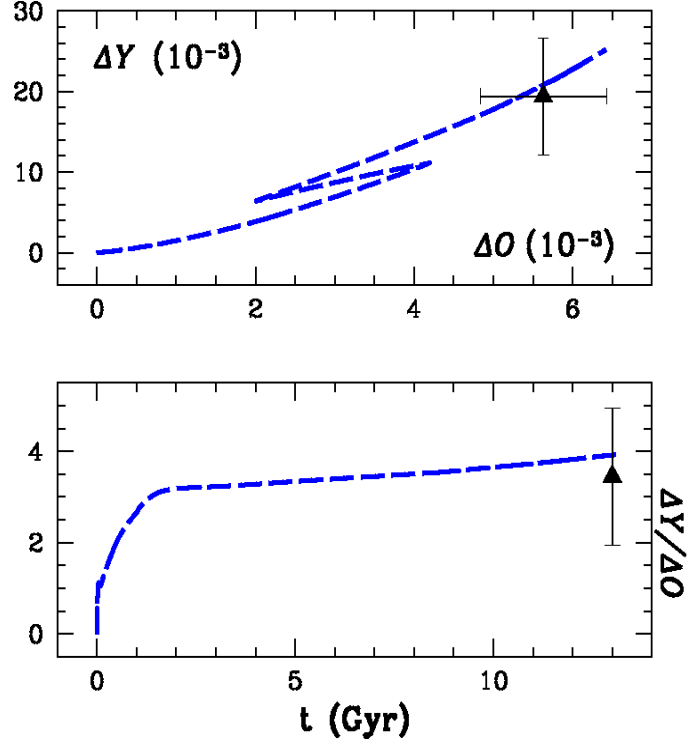

Fig. 12. Results of CEM at r = 15.0 kpc (dashed-blue line). Top panel: $\Delta Y$ versus $\Delta O$ evolution. Bottom panel: $\Delta Y/\Delta O$ evolution. Observational data at r = 14.7 kpc: oxygen value obtained from the O/H gradient (see equation 4 in Section 2.3), and helium value derived in Section 5.

There are two H II regions of M31 with the best He/H determinations in the literature, objects 25 and 26 by Zurita & Bresolin (2012), both regions at a galactocentric distance of 14.7 kpc. These authors derived $He^+/H^+$ values of 0.086±0.004 and 0.088±0.005 for regions 25 and 26 respectively, based on the He atomic data by Porter et al. (2005, 2007).

Based on the He I and H I emission lines by Zurita & Bresolin (2012), we use the program by Peimbert et al. (2012) with the He I atomic data by Porter et al. (2013) and derive values of $He^+/H^+$ of 0.0792 and 0.0839 for objects 25 and 26, respectively. To correct for the presence of neutral helium we assume that $He^0/He = S^+/S$, and based on the $S^+/S$ ratios presented by Zurita & Bresolin (2012), obtain values of He/H of 0.0876±0.0050 and 0.0981±0.0050 for regions 25 and 26. Therefore, for r = 14.7 kpc, we adopt a value of He/H=0.0928±0.0035, the average of both determinations.

For a galactocentric distance of 14.7 kpc and based on the O/H gradient (see equation 4) we obtain that $12 + \log(O/H) = 8.70 \pm 0.06$. Under the assumption that O amounts to 42% of the total Z value (Peimbert et al. 2007b) it follows that $X = 0.7195 \pm 0.0074$, $Y = 0.2671 \pm 0.0073$ and $Z = 0.0134 \pm 0.0020$.

From the X, Y and Z values we obtain an O by mass of $0.00563 \pm 0.0008$, and a $\Delta Y = 0.0194 \pm 0.0079$,

consequently ΔY/ΔO = 3.45 ± 1.50. As in our previous works, we have adopted the Yp value by Peimbert et al. (2007a) of 0.2477 ± 0.0029. This Yp value is in very good agreement with the Planck satellite prediction of Yp, in standard Big Bang nucleosynthesis, that amounts to 0.2473 ± 0.0003 (Ade et al. 2013).

## 6. CONCLUSIONS

We reach the following conclusions:

(1) The O/H abundance ratios from A-F supergiant stars agree with the abundancesderived from H II regions only if the gaseous abundances determined from recombination lines are corrected by dust depletion. Moreover, the stellar O/H values agree with the O/H gaseous values derived from collisional excitation lines only if they are increased by the effect of temperature variations and by the fraction of O atoms embedded in dust grains.

(2) After correcting the O/H abundances from H II regions by $t^2$ and dust depletion, the O/H gradient in the 7-18 kpc region is well represented by:

$12 + \log(O/H) = 8.73 \pm 0.06) - (r-12)(0.013 \pm 0.006)$.

(3) An upper mass in the IMF for M31 equal to 45$\odot$ is required to reproduce the absolute value of the observed O/H gradient.\(4) Our model is in good agreement with the O/H, C/H, N/H, Mg/H, Si/H, S/H, Ar/H, Cr/H, Fe/H, and $Z$ gradients.

(5) The original yields of Ne, Al, Sc, Ti, Mn and Ni computed by Kobayashi et al. (2006) have to be multiplied by 0.35, 2.5, 10, 5, 8 and 5, in order to reproduce the absolute value of the Ne/H, Al/H, Sc/H, Ti/H, Mn/H and Ni/H gradients, respectively.

(6) The ΔY/ΔO value determined from observations is in agreement with the model.

(7) M31 has been formed according to a pronounced inside-out scenario.

(8) A star formation efficiency ν(r), variable in space and constant in time for most of the evolution is necessary to reproduce the current profile of the gas mass density for the whole galactic disk, for most of evolution.\

(9) A considerable reduction of the ν value, for t> 12.8 Gyr and for r> 12 kpc is inferred to explain the important decrease of the SFR observed in the outer regions.

(10) Each predicted Xi/H gradient shows three slopes during the whole evolution, due to the three r-behaviours of ν. These slopes flatten at different rates, due to the combination of ν(r) and the infall rate that depends on r and t, according to the inside-out scenario. For the inner and outer part the current gradients are slightly positive, while in the middle part the slope is slightly negative. More and better determinations of chemical abundances are required to confirm these results.

# APPENDIX: SOME UNSUCCESFUL MODELS

In this appendix we show four of the hundreds of CEMs built for this study. In particular, we present those models that reproduce some, but not all, of the main observational constraints for all galactocentric distances.

The tested models are products of modifying three main ingredients of the CEMs:i) the inside-out efficiency of the galactic formation, varying τdisk(r); ii) the efficiency of the star formation rate, varying ν(r), but constant in time; and iii) the number of massive stars, varying the mup of the adopted IMF.

In each tested model, we assume the same ingredients of our best model, except the modified specific parameter. It is important to remind that our best model, presented in sections 3 and 4, is characterized by a pronounced inside-out scenario (τdisk = 0.3r - 0.2), an u-shaped ν(r) for t< 12.8 Gyr, an exponential-shaped ν(r) t> 12.8 Gyr (see Figure 3), and mup=45 M⊙.

In the figures below, we compare our tested models with the radial distributions of Mgas, O/H, and SFR observed in the M31 disk.

*A less pronounced inside-out scenario*

In Figure 13, we present the CEM that assumes a τdisk = 1.0r – 2.0. This model predicts Mgas values higher than observed, mainly for r< 12 kpc, this fact produces a higher SFR in disagreement with observations. The predicted O/H gradient fits, within errors smaller than 2 σ, most of the O/H values. At recent time, the slower decrease of the infall with time, implies a higher amount of falling material that produces higher Mgas values that enter in conflict with the observations.

*Independent-time efficiency of the SFR*

U-shaped ν(r)

In Figure 14, we present the CEM that considers the u-shaped ν( r) shown in the upper panel of Figure 3, during the whole evolution (0 <t(Gyr)< 13.0). This model predicts Mgas(r) and O/H(r) that adjust very well the observations, but the predicted SFR for r> 12 kpc is slightly higher than the observations. As mentioned in the section 4, the need to vary ν at recent time is to reproduce the current SFR in the outer disk.

Exponential-shaped ν(r)

In Figure 15, we present the CEM that takes into account the exponential-shap ν(r) shown in the bottom panel of Figure 3, during the whole evolution for 0 <t(Gyr)< 13.0.For r> 12 kpc, this model predicts Mgas values higher and O/H values lower than observed, because the efficiency of the SFR is very slow, producing a smaller number of stars than our best model. The predicted SFR for r> 12 kpc is higher than observed, due to the increased Mgas values.

## A Higher value of mup

In Figure 16, we present the CEM with an mup = 80M⊙. As expected, this model predicts Mgas and SFR values very similar to those obtained by the best model, but the absolute value of the O/H distribution is higher than most of the observed O/H values.

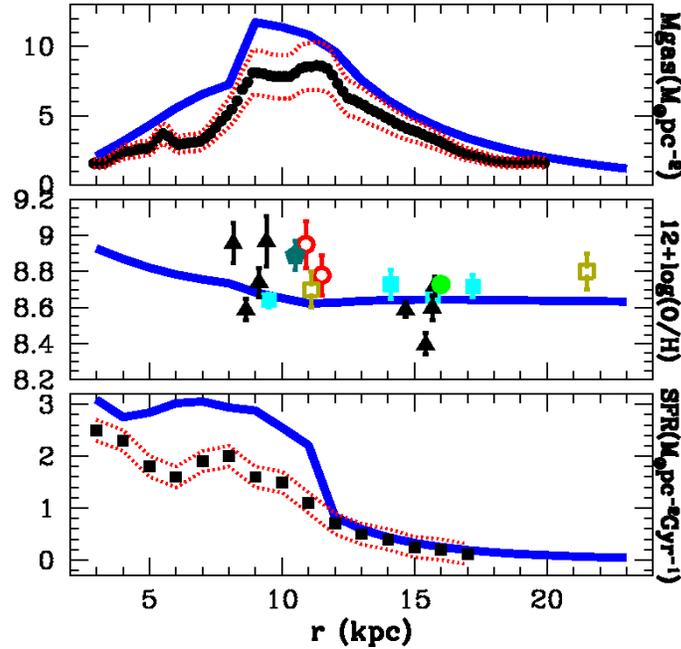

Fig. 13. Radial distributions of the gas mass, O/H, and star formation rate (top, middle, and bottom panels). Results at the present time of a chemical evolution model that considers a less pronounced inside-out scenario of (solid blue lines). Observational data as presented in Figs. 1 and 2.

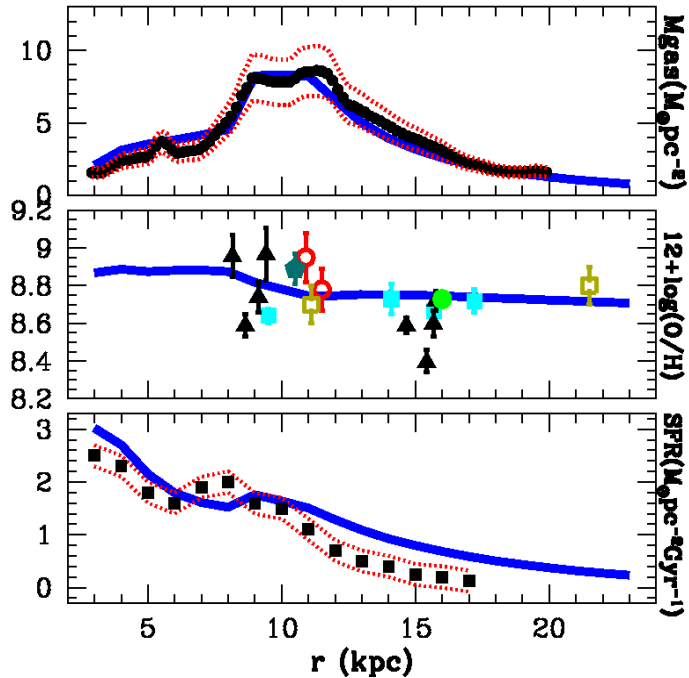

Fig. 14. CEM that considers the u-shaped v(r) of the star formation rate during the whole evolution ($0 < t(Gyr) < 13.0$). This v(r) is shown in Fig. 3 (upper panel). Panels and data as in Fig 13.

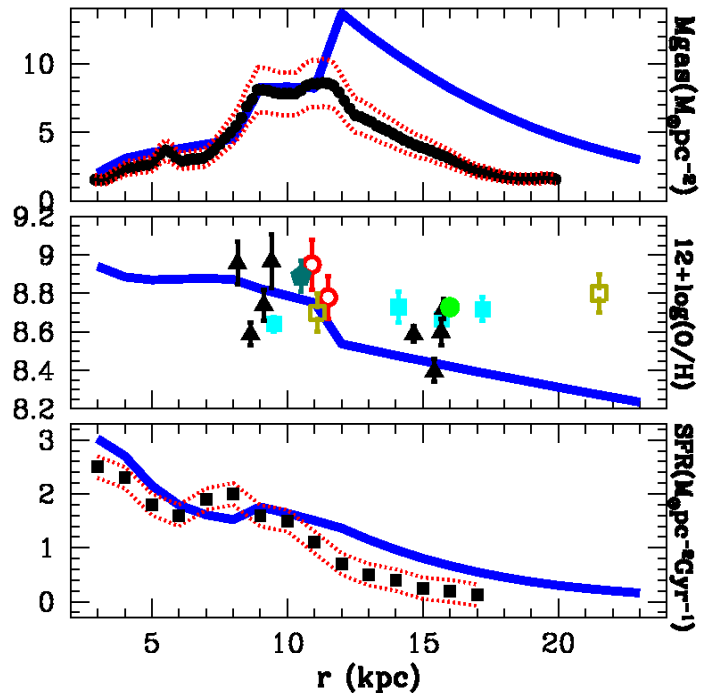

Fig. 15. CEM that considers the exponential-shaped v(r) of the star formation rate during the whole evolution (0 < t(Gyr)< 13.0). This v(r) is shown in Fig. 3 (bottom panel). Panels and data as in Fig 13.

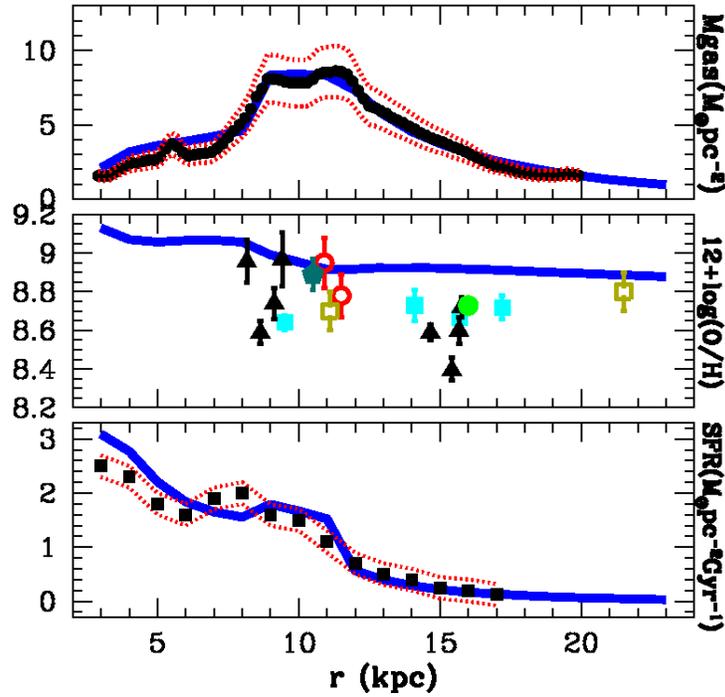

Fig. 16. CEM that considers 80M⊙ as the upper mass of the initial mass function. Panels and data as in Fig 13.

## ACKNOWLEDGMENTS

We thank the referee for a careful review of the manuscript and several useful suggestions, that contributed to improve this paper. We are grateful to Antonio Peimbert for many helpful discussions, to Fatemeh Tabatabaei for providing us with the updated data of the surface density of the atomic and molecular hydrogen and the financial support provided by CONACyT of Mexico (grant 129753). F. Robles-Valdez is grateful to CONACyT, México, for a doctoral grant. L. C. thanks the financial support provided by the Ministry of Science and Innovation of the Kingdom of Spain (grants AYA2010-16717 and AYA2011-22614).